\documentclass[a4paper,prb,aps,twocolumn]{revtex4}

\usepackage{amsmath}
\usepackage{amssymb}
\def\sigmaxc{\stackrel{\leftrightarrow}{\sigma}}

\begin{document}

\title{Foundations of stochastic time-dependent current-density functional
   theory for open quantum systems: Potential pitfalls and rigorous
   results}

\author{Roberto D'Agosta}\email{roberto.dagosta@ehu.es}
\affiliation{ETSF Scientific Development Center,
Departamento de F\'isica de Materiales, Universidad del Pa\'is Vasco, E-20018
San Sebasti\'an, Spain}
\affiliation{IKERBASQUE, Basque Foundation for Science, E-48011,
Bilbao, Spain}

\author{Massimiliano Di Ventra}\email{diventra@physics.ucsd.edu}
\affiliation{Department of Physics, University of California -
San Diego, 92093 La Jolla, USA}

\begin{abstract}
We clarify some misunderstandings on the time-dependent
current density functional theory for open quantum systems we have recently
introduced [M. Di Ventra and R. D'Agosta, Phys. Rev. Lett. {\bf 98}, 226403
(2007)]. We also show that some of the recent formulations attempting to
improve on this theory suffer from some inconsistencies, especially in
establishing the mapping between the external potential and the quantities of
interest. We offer a general argument about this mapping, that applies to any density functional theory, showing that it must
fulfil certain ``dimensionality" requirements.
\end{abstract}
\maketitle

\section{Introduction}
Experience teaches us that no physical system can be considered as completely
closed, in the sense that some degrees of freedom are either too many to
consider explicitly, or their microscopic features are of no interest, and
only their macroscopic thermodynamic properties are of importance. These
degrees of freedom are then treated as baths or reservoirs of the system of
interest, making the latter ``open''. Nonetheless, the theory of open quantum
systems is relatively new, starting with the pioneering works by Zwanzig and
Nakajima,\cite{Zwanzig1960,Nakajima1958} who worked out the theory in terms of
the statistical operator (density matrix), and more recently with the
formulation of the theory of open quantum system in terms of a state vector
(``wavefunction"). \cite{Gisin1981,Strunz1996}

In parallel, another theory, Density Functional Theory (DFT), developed since
the first papers by Kohn and co-workers, \cite{Hohenberg1964,Kohn1965} has
changed the way we think in terms of many-body quantum mechanical systems.
Indeed, the successes of DFT are countless, and the theory has now become a
standard for many calculations of electronic structure, electron transport,
materials properties, and chemical
reactions.\cite{Dreizler1990,Trickey1990,Nalewajski1996,Giulianivignale,DiVentra2008,Marques2006}

The attempt to bring together the two theories appears therefore completely
natural.\cite{Burke2005,DiVentra2007,DAgosta2008a,Appel2009,Yuen-Zhou2009a,Yuen-Zhou2010,Biele2012} In particular, the present authors have proved a theorem
that establishes a one-to-one correspondence between a vector potential acting
on an open quantum system and its current
density,\cite{DiVentra2007,DAgosta2008a} a theorem that was later extended to
the correlated motion between electrons and (in principle quantum)
ions.\cite{Appel2009,Appel2011} This theory, which we named stochastic
time-dependent current-DFT (STDCDFT), then allows for the solution of a
many-body open quantum system with effective single-particle equations, a
significant computational simplification.

Although still in its infancy, STDCDFT for open quantum systems has received a
certain amount of attention, especially for its promise to expand the field to
systems previously inaccessible within standard DFT. However, there is also
some confusion regarding its foundations which is mostly evident in recent
developments, for instance a new result that shows how in principle we can
mimic the dynamics of an open quantum system by reverting to the analysis of a
closed quantum system made of non-interacting
particles.\cite{Yuen-Zhou2009a,Yuen-Zhou2010} This result would further
dramatically reduce the computational cost for evaluating the dynamics of the
open quantum system.

The reason for this reduction is simple. In general, the open quantum system
could be described in two ways. On the one hand, we can use the many-body
statistical operator, or the single-particle reduced density matrix. However,
for large systems, the number of degrees of freedom, i.e., the dimensionality
of the operator we need to consider, scales badly although some schemes can be
devised to cure this scaling.\cite{Pershin2008a} On the other hand, a
formulation in terms of a stochastic state vector is computationally cheaper
than a density matrix one, but nonetheless it requires the consideration of a
statistical ensemble of replicas of the system. Larger the ensemble, finer will
be the physical information we can extract from the dynamics. Instead, if we
could describe the dynamics of the open system with a closed non-interacting
system, we would have reduced a complex and expensive problem to a size similar
to that we usually tackle with our present computational means.

Obviously,--a situation common in DFT--we cannot expect to have access to the
complete physical information about the open quantum system. We must accept
that our non-interacting {\it doppelganger} will be able to describe only some
quantity of choice, normally either the single-particle density or the
single-particle current density. In principle, any other physical observable
can be extracted provided we know its expression in terms of these
``fundamental'' quantities. However, in practice, we rarely know how to write
such observables in terms of the density or the current density, thus limiting
the range of physical quantities that can be actually computed. Our goal for this paper is to both clarify some of the theoretical foundations
of STDCDFT that do not seem to have been appreciated in the literature, as well
as to show that some of the recent attempts to extend this theory
\cite{Yuen-Zhou2009a,Yuen-Zhou2010} suffer from inconsistencies, in particular
those related to the mapping between external potentials and the basic
quantities of any density functional theory.

\section{On the mapping between vector potential and densities}
Let us then start by noting that the choice about which quantity we want to
base our theory on is usually made \emph{a priori}. For this reason we usually
talk about DFT or Current-DFT.\cite{Vignale1988,Giulianivignale} For closed
systems, Current-DFT is regarded as a more complete theory than DFT since, via
the continuity equation, one can calculate the single particle density starting
from the single-particle current density. The situation is instead more complex
in open quantum systems since, as we are going to show in the following,
\emph{prima facie} the continuity equation would seem not to uniquely determine
the single-particle density, given the single-particle current density.

Let us then begin by recalling that in the theory of open quantum systems, the
equation of motion for the ensemble-averaged particle density, $n(r,t)$, is
given by (see, e.g., Ref. \onlinecite{Frensley1990})
\begin{equation}
	\partial_t n(r,t)=-\vec{\nabla}\cdot \vec{j}(r,t)+\mathcal{F}_B(r,t)
	\label{continuity}
\end{equation}
where $\mathcal{F}_{B}(r,t)$ describes the density modulation induced by the
presence of the bath(s) and $\vec{j}(r,t)$ is the ensemble-averaged current
density. Notice that in this theory we do not allow for the system to exchange
particles with the environment. Eq. (\ref{continuity}) can be obtained from the
master equation of motion for the density matrix, $\hat\rho$ ($\hbar=1$)
\begin{equation}
	\partial_t \hat \rho(t)=-i\left[ \hat
H(t),\hat \rho(t)\right] +\int_{t_0}^t dt' K(t,t')\hat \rho(t').
\label{master}
\end{equation}
Here, $K(t,t')$ is a memory kernel that describes the action of the bath(s) on
the system--provided it is chosen to preserve the positivity of the density
matrix at any given time--
\cite{Zwanzig1960,Nakajima1958,Gardiner2000,vanKampen,Biele2012} and $\hat
H(t)$ is the Hamiltonian of the system that evolves under the action of an
external vector potential (possibly time dependent), $\vec A(r,t)$, and in the
presence of a particle-particle interaction operator $\hat W$,
\begin{equation}
	\hat H(t)=\frac{1}{2m}\sum_i\left(\hat{\vec{p}_i}+e\vec A(\hat r_i,t)\right)^2+\hat W
	\label{h_manybody}
\end{equation}
We work here in a gauge where the scalar potential has been set to zero at all
times.

In general, little is known about the term $\mathcal F_B$ appearing in the
continuity equation (\ref{continuity}). Although not strictly necessary for the
considerations we make later, it would be useful if $\mathcal F_B$ could be
written as the divergence of some current. This would immediately imply the
conservation of the total number of particles. Recently, Gebauer and Car have
shown that this is indeed the case in the Markov approximation for certain
couplings between the system and the
environment.\cite{Gebauer2004a,Gebauer2005} More recently a formal condition on
current conservation and its use to test the validity of the approximations
made to arrive at the master equation has been proposed.\cite{Salmilehto2012}
If $\mathcal F_B$ is the divergence of some current--this extra current being
generated by the exchange of momentum and energy with the external
environment--then the continuity equation assumes the usual physical meaning
and particle number conservation is guaranteed. Moreover, one could assume that
\begin{equation}
	\mathcal{F}_B=-\nabla\cdot \left(\frac{e\vec C(r,t)}{m}n(r,t)\right)\,,
	\label{leakage}
\end{equation}
where without loss of generality we have written the extra current as $e
n(r,t)\vec C(r,t)/m$. If this were the case, the total current $\vec
j_T(r,t)=\vec j(r,t)+e n(r,t)\vec C(r,t)/m$ fulfills the standard continuity
equation,
\begin{equation}
	\partial_t n(r,t)=-\nabla \cdot \vec j_T(r,t).
\end{equation}
In the total current, $\vec C(r,t)$ appears then to play the role of an extra
vector potential, effectively inducing a diamagnetic current in the system,
$en(r,t)\vec C(r,t)/m$. Interestingly, in Ref.~\onlinecite{Yuen-Zhou2009a} $\vec
C(r,t)$ has been dubbed ``leakage potential'', although we would like to point
out that its definition from Eq. (\ref{leakage}) essentially prevents any
leakage, i.e. loss of particles, from the system.

In standard TDCDFT for closed quantum systems, where the extra term in Eq.
(\ref{continuity}) is not present, i.e., $\mathcal{F}_B(r,t)=0$, the continuity
equation establishes an important link between the single-particle current
density $\vec j(r,t)$ and the single-particle density $n(r,t)$. Indeed, this
equation allows for expressing the latter in terms of the former via a simple
time integration. This seems a trivial point, but let us clearly state it: The
continuity equation, when $\mathcal{F}_B(r,t)=0$ can be used together with the
initial conditions to {\it uniquely} determine the single-particle density
given the single-particle current density. The basic theorem of TDCDFT
therefore establishes a one-to-one mapping between the current density and the
vector potential applied to the system, once the initial conditions are
provided.\cite{Vignale2004,Giulianivignale,Marques2006} In fact, whenever one
discusses the uniqueness of the mapping between the single-particle density and
single-particle current density with the external vector potential, it is
implicitly understood that the particle density is obtained from the current
density via the continuity equation. If expressed in terms of the set of
densities $\left \{ \vec j(r,t),n(r,t) \right \}$, the mapping between this set and the vector
potential would be redundant, in the sense that one of the variables--usually
$n(r,t)$--can be determined by $\vec j(r,t)$. As an exercise, in the following subsection we will detail the proof of the standard TDCDFT theorem for closed system, without the requirement that the single particle density is the same in the Kohn-Sham and in the real system.

The situation becomes a bit trickier whenever $\mathcal{F}_B(r,t)\not = 0$. In
this case one needs the expressions of both $\vec j(r,t)$ and
$\mathcal{F}_B(r,t)$ to be able to determine the single-particle density.
Indeed, written in this way the problem corresponds to the solution of a linear
differential equation with non-constant coefficients. Then, if one knows $\vec
j(r,t)$ and $\mathcal F_B(r,t)$ we just need to perform a simple time
integration to obtain $n(r,t)$. From a TDCDFT formulation of the theory then,
as it has been shown,\cite{DiVentra2007,DAgosta2008a} one can build a
one-to-one mapping between the (average) current density and the external
vector potential, provided the initial condition and the bath operators are
fixed. In Ref.\onlinecite{DiVentra2007} the thesis of the theorem is that there
is a one-to-one mapping between the current density and the vector potential,
but in the proof we have implicitly assumed that the particle density could be
determined from the continuity equation. In Ref.~\onlinecite{DAgosta2008a} we
have clarified the role of the continuity equation and assumed that the single
particle density is determined uniquely by the current density -- unfortunately
for open quantum system we do not have a proof of this statement. After
that, we have formulated the theorem by including the single particle density
in the mapping. In this respect, our mapping could appear identical to the one
presented in Ref.~\onlinecite{Yuen-Zhou2009a}, with the distinction however
that for us the single particle density is fixed uniquely by the current
density.

Here, we stress that the inclusion in the mapping of the density as a variable
along with the current density--as for example it has been tried in
Ref.~\onlinecite{Yuen-Zhou2009a}--is not just redundant, but simply incorrect.
Indeed, we will show in the following that the mapping between the set of
current and particle densities and the {\it vector} potential is not unique.
Our argument can be extended to other cases. For example, along the same lines
we can prove that the mapping between the current density and the {\it scalar}
potential cannot be one-to-one. One of us (RD) has already presented another
proof of the same result, in particular showing that the current density is in
general non-V-representable.\cite{DAgosta2005a} This, therefore, precludes the
access to all components of the current density by standard TDDFT. Our
following argument also sets a necessary condition for building any density
functional theory, in terms of establishing a one-to-one correspondence between
two physical quantities. This condition states that the two quantities must
have the same ``dimensions'': a vector with a vector, a scalar with a scalar,
and so on. Any other mapping will be either redundant or wrong.

\subsection{A theorem of standard TDCDFT}\label{standardtdcdft} Before
proceeding we would like to extend the Vignale's proof of the fundamental
theorem of TDCDFT\cite{Vignale2004}. In this extension, we explicitly assume
that the continuity equation is not valid, and therefore that the single
particle densities in the real and Kohn-Sham system do not coincide. In this
section, with a little abuse of notation, we will define $n(r,t)$ and $\vec
j(r,t)$ the single particle density and current density, while elsewhere these
define the corresponding ensemble-average quantities. We will prove that:
\emph{Theorem (mapping between vector potential and current density)} Consider a many-particle system described by the time-dependent Hamiltonian
\begin{equation}
	\hat H(t) =\sum_i\left[\frac{1}{2m}\left( \hat p_i+\vec A(\hat r_i,t)\right)^2\right]+\sum_{i<j}U(\hat r_i-\hat r_j)
\end{equation}
where $\vec A(\hat r,t)$ is a given external vector potential, which is an analytic function of time
in a neighborhood of $t=0$, and $U(\hat r_i-\hat r_j)$ is a translationally invariant two-particle interaction. Let
$\vec j(r,t)$ be the current density that evolves under $\hat H$ from a given initial state $|\Psi(0)\rangle$. Then, under reasonable assumptions, the same current density can be obtained from another
many-particle system, with Hamiltonian
\begin{equation}
\hat H'(t) =\sum_i\left[\frac{1}{2m}\left( \hat p_i+\vec A'(\hat r_i,t)\right)^2\right]+\sum_{i<j}U'(\hat r_i-\hat r_j),
\end{equation}
starting from an initial state $|\Psi'(0)\rangle$ which yields the same
current density as $|\Psi(0)\rangle$ at time $t = 0$. The potential $\vec A'(r,t)$, is uniquely determined by $\vec A(r,t)$, $|\Psi(0)\rangle$, and $|\Psi'(0)\rangle$, up to gauge transformations of the form
\begin{eqnarray}
	\vec A'(r,t)\rightarrow \vec A'(r,t)+\nabla \Lambda(r,t),
\end{eqnarray}
where $\Lambda$ is an arbitrary regular function of $r$ and $t$, which
satisfies the initial condition $\Lambda(t=0)=0$.

{\it Proof:}
The proof of this theorem follows in the same footsteps of similar proofs
already present in the literature for TDDFT and
TDCDFT.\cite{Runge1984,vanLeeuwen1999, Vignale2004} More advanced proofs have
been recently suggested especially for the basic theorem of TDDFT
\cite{Ruggenthaler2011} and for TDCDFT on a lattice \cite{Tokatly2011}. One of
the advantages of these novel proofs is that they remove the analyticity
requirement around the initial time. Although research in this direction is
currently under way \footnote{P. de Boeji, private communication.}, at the moment
there is not a similar proof for TDCDFT so we will use the standard way to
prove this theorem. We assume that the same current density ${\vec j}(r,t)$ is
also obtained from another many-particle system with Hamiltonian
\begin{equation}
\hat H'(t)=\sum_{i} \frac{\left[\hat p_i+e
    {\vec A}'(\hat r_i,t)\right]^2}{2m}+\frac12 \sum_{i\not = j}U'\left(\hat r_i-\hat r_j\right), \label{h1}
\end{equation}
evolving   from   an  initial   state   $|\Psi'_0\rangle$   and  following   the
time-dependent Schr\"{o}dinger equation. $|\Psi'_0\rangle$  gives, in the primed
system, the same initial current density as in the unprimed system.

The core of the demonstration is as follows: We write the equations of motion
for ${\vec j}(r,t)$ determined by both ${\vec A}(r,t)$ and ${\vec A}'(r,t)$,
and compare these two equations to obtain an equation for ${\vec A}'(r,t)$ in
terms of ${\vec A}(r,t)$. We then prove that ${\vec A}'(r,t)$ is completely
determined by the initial condition via a series expansion in time about $t=0$.
Finally, if the two systems coincide then the unique solution is ${\vec
A}(r,t)={\vec A}'(r,t)$ up to a gauge transformation.

The equation of motion for the current density is
easily obtained from the equation of motion of the
current density operator. A standard derivation leads us to
\begin{eqnarray}
\partial_t {\vec j}(r,t)&=&\frac{n(r,t)}{m} \partial_t {\vec A}(r,t) -\frac{{\vec j}(r,t)}{m}\times\left[\nabla \times {\vec A}(r,t)\right]\nonumber\\
&&+\frac{{\vec {\mathcal G}}(r,t)}{m}\label{currenteq}
\end{eqnarray}
where we have defined the vector of the internal forces
\begin{equation} \label{forces}
{\vec {\mathcal G}}(r,t)=-\left\langle\sum_{i\not = j}\delta(r-\hat r_i)
 \nabla_j U\left(\hat r_i-\hat r_j\right)\right\rangle+m \left\langle\nabla
\cdot \hat \sigmaxc(r,t)\right\rangle
\end{equation}
with the stress tensor $\hat \sigmaxc(r,t)$ given by
\begin{equation}
\hat \sigma_{i,j}(r,t)=-\frac{1}{4}\sum_k \{\hat v_i,\{\hat
v_j,\delta(r-\hat r_k)\}\}.
\label{stresstensor}
\end{equation}
The first two terms on the right hand side of Eq.~(\ref{currenteq}) describe the
effect of the applied electromagnetic field on the dynamics of the
many-particle system while the third is due to particle-particle
interactions.

Equations similar to Eqs.~(\ref{currenteq}) -- (\ref{stresstensor}) can now be
written for the system with the vector potential ${\vec A}'(r,t)$. A similar
force term $\mathcal{G}'$ appears in these new equations and differs from the
same forces in the unprimed system, since the initial state, the external
vector potentials and the velocity $\hat v$ are different. By assumption, the current density is the same in the two systems, thus
\begin{eqnarray}
\partial_t {\vec j}(r,t)&=&\frac{n'(r,t)}{m} \partial_t {\vec A}'(r,t)-\frac{{\vec j}(r,t)}{m}\times\left[\nabla \times {\vec A}'(r,t)\right]\nonumber\\
&&+\vec {\mathcal G}'(r,t)\label{currentqp}.
\end{eqnarray}
In Eq. (\ref{currentqp}) we have explicitly taken into account the different
single particle density in the primed system. Taking the difference of
Eqs.~(\ref{currenteq}) and (\ref{currentqp}) we arrive at
\begin{eqnarray}
n'(r,t) \partial_t \vec A'(r,t) &=&n(r,t) \partial_t \vec A(r,t)\nonumber\\
&&-\vec j (r,t) \times\left[\nabla \times \left({\vec A}(r,t)-{\vec A}'(r,t)\right)\right]\nonumber\\
&&+m\vec {\mathcal G}(r,t)-m\vec {\mathcal G}'(r,t). \label{difference}
\end{eqnarray}

We now need to prove that Eq.~(\ref{difference}) admits only one solution,
i.e., ${\vec A}'(r,t)$ is completely determined by the dynamics of the current
density. To this end we expand Eq.~(\ref{difference}) in series about $t=0$ and
obtain an equation for the $l$-th derivative of the vector potential $\vec
A'(r,t)$. That one can expand this equation in a time series about $t=0$
follows immediately from the analyticity of the vector potential and the
properties of the Sch\"odinger equation. We thus arrive at the equation
\begin{widetext}
\begin{equation}
\begin{split}
n'_0(r)(l+1)\vec A'_{l+1}(r)=&n_0(r)(l+1)\vec A_{l+1}(r)-\sum_{k=0}^{l-1}(k+1)\left(n'_{l-k}(r)\vec A'_{k+1}(r)-n_{l-k}(r)\vec A_{k+1}(r)\right)\\
&+m\vec {\mathcal G}_l(r)-m\vec {\mathcal G}'_l(r)+\sum_{k=0}^l {j_{l-k}(r)}\times \left[\nabla\times
\left(\vec A'_{k}(r)-\vec A_{k}(r)\right)\right] \label{expansion}
\end{split}
\end{equation}
where, given an arbitrary function of time $f(r,t)$, we have
defined the series expansion
$f_l(r)\equiv \frac{1}{l!}\left.\frac{\partial^l f({\bf
r},t)}{\partial t^l}\right|_{t=0}.$
\end{widetext}
It is important to point out here that the initial density $n'_0(r)=n'(r,t=0)$
is determined solely by the initial conditions of the primed system. This
quantity is therefore accessible to us, since we assume that the initial
conditions are assigned both in the primed and unprimed systems. We are now
left to prove that the right hand side of Eq.~(\ref{expansion}) does not
contain any term $A'_{l+1}(r)$. This follows directly from the structure of the
time-dependent Sch\"odinger equation. Indeed, this implies that the $l$-th time
derivative of any operator can be expressed in terms of its derivatives of
order $k<l$, and time derivatives of the Hamiltonian of order $k<l$. The time
derivatives of the Hamiltonian do contain time derivatives of the vector
potential ${\vec A}(r,t)$, but always of order $k<l$. Then on the right hand
side of Eq.~(\ref{expansion}) no time derivative of order $l+1$ appears.
Equation~(\ref{expansion}) can be thus viewed as a recursive relation for the
time derivatives of the vector potential $\vec A'(r,t)$. To complete the
recursion procedure we only need to assign the initial value ${\vec
A}'(r,t=0)$. Since in the unprimed and primed systems the current densities
are, by hypothesis, equal, the initial condition is simply given by
$n'(r,t=0)A'_0(r)=n(r,t=0)A_0(r)+\langle \Psi_0|\overline{\hat j_p
(r,t=0)}|\Psi_0\rangle -\langle \Psi'_0|\overline{\hat j_p (r,t=0)}|\Psi'_0
\rangle$, where $\hat j_p(r)=(1/2m)\sum_i \{\hat p_i,\delta(r-\hat r_i)\}$ is
the paramagnetic current density operator.

The same considerations as in Ref.~\onlinecite{Vignale2004} about the
finiteness of the convergence radius of the time series~(\ref{expansion}) apply
to our case as well. To finalize our proof, we consider the case in which
$U=U'$ and $|\Psi_0\rangle=|\Psi'_0\rangle$. If this holds, $ A'_0(r)=A_0(r)$.
Then the recursion relation admits the unique solution $\vec A'_l(r)\equiv \vec
A_l(r)$ for any $l$, and at any instant of time $t$ we have $A(r,t)=A'(r,t)$
(still up to a gauge transformation). On the other hand, we can assume that
$U'=0$, so that the primed system is made of non-interacting particles, {\it
i.e.} the Kohn-Sham system.

This theorem shows that the standard TDCDFT can be reformulated in terms of
{\it only the current density}. The presence of the continuity equation makes
the proof more easy and natural. We can therefore conclude that {\it any}
theorem of TDCDFT that includes the current and particle density in the map
with the vector potential creates a redundant mapping since the continuity
equation uniquely determines the single-particle density.

The proof of the above theorem also gives us an important tool. Since proofs of the theorem of STDCDFT and any theorem of a time-dependent current-density functional theory for open systems, are essentially based on this very same proof, we can extend it immediately to the theory of open quantum system. This shows that there is a one-to-one mapping between the ensemble-averaged current density and the external vector potential.

\subsection{A simple counterexample}
In fact, we can provide a simple counter argument to the existence of a mapping
between the vector potential and the set of functions given by the particle and
current densities. For this mapping to be non-redundant we need to assume for
example that the particle density is not determined by the sole knowledge of
the current density. So in the following we are assuming that it is not
possible to establish a connection between the particle density and the current
density. This would be the case in the DFT for open quantum system if the
function $F_B$ cannot be expressed as a functional of the current and particle
densities or, if such a functional expression exists, the now-non-linear
continuity equation cannot be solved to give the particle density.

Let us therefore consider the extremely simple case in which the response
functions of the system are constant in time and in space, and that the (full)
response is linear. We can easily generalize this counterexample to the case of
response functions that are local in both space and time. Moreover, since the Fourier
transform creates a unique mapping between function in real and momentum space,
this counterexample can be generalized to the response to a monochromatic
external vector potential, thus making it applicable to the case of a uniform
free electron gas subject to an external perturbation.\cite{Giulianivignale,DAgosta2005a}
For this system we can write the following
response equations (for simplicity, in the next few equations we drop the
spatial and temporal dependence of $j$ and $n$)
\begin{equation}
	\left(
	\begin{array}{c}
		j_x\\ j_y \\ j_z \\n
	\end{array}
	\right)
	=
	\hat\Gamma
	\left(
	\begin{array}{c}
		A_x\\ A_y \\ A_z
	\end{array}
	\right)
	\label{linearresponse}
\end{equation}
where $\Gamma$ is a $4\times3$ matrix of constant coefficients of the system
response. The existence of a one-to-one mapping between the particle and
current densities and the vector potential would then suggest that we can write
a similar equation, starting from the particle and current densities
\begin{equation}
	\left(
	\begin{array}{c}
		A_x\\ A_y \\ A_z
	\end{array}
	\right)
	=
	\hat\Gamma'
	\left(
	\begin{array}{c}
		j_x\\ j_y \\ j_z \\ n
	\end{array}
	\right)
	\label{inverselinearresponse}
\end{equation}
where now $\Gamma'$ is a $3\times4$ matrix. Combining these results it follows
that
\begin{equation}
	\hat \Gamma \hat \Gamma'=1_4;~ \hat \Gamma'\hat \Gamma=1_3,
\end{equation}
where $1_N$ is the $N\times N$ identity matrix.

We can now prove that this mapping is not one-to-one. Since Eq. (\ref{inverselinearresponse}) is a system of 3
equations in 4 variables, we can always find two distinct sets of current and
particle densities providing the same vector potential. Therefore an infinite
number of solutions can be found. To show this we proceed in this way: Assume
that $\vec j_0, n_0$ are a set of current and particle densities, and that $\vec
A_0$ is the resulting vector potential obtained from
(\ref{inverselinearresponse}). Now let us consider the particle density
$n=n_0+\Delta n$. We want to show that we can find another current density
$j=j_0+\Delta j$ which together with $n$ gives, via
(\ref{inverselinearresponse}), the same vector potential. Since $\Delta n$ is
arbitrary we can choose as initial condition $\Delta n(r,t=0)=0$, therefore the
two sets of current and particle densities do satisfy the same initial
conditions. To prove our point, it is enough to prove that the system of
equations
\begin{equation}
	\hat\Gamma'
	\left(
	\begin{array}{c}
		j_{0,x}\\ j_{0,y} \\ j_{0,z} \\ n_0
	\end{array}
	\right)=
	\hat\Gamma'
	\left(
	\begin{array}{c}
		j_x\\ j_y \\ j_z \\ n
	\end{array}
	\right),
\end{equation}
or the equivalent
\begin{equation}
	\hat\Gamma'
	\left(
	\begin{array}{c}
		\Delta j_x\\ \Delta j_y \\ \Delta j_z \\ \Delta n
	\end{array}
	\right)=0
	\label{homogeneous}
\end{equation}
admits at least one solution. This is again trivially true, since this is a
system of 3 equations in 4 variables. We can now assume the density $\Delta n$
assigned and find, if the 3x3 submatrix of $\Gamma'$ is invertible, the
corresponding current density that satisfies Eq. (\ref{homogeneous}). If that
submatrix of $\Gamma'$ is not invertible, then we will be able to find more
than one solution for $\Delta j$, thus reinforcing our statement. The solution
found in this way can also be made to fulfill the initial condition $\Delta
\vec j(r,t=0)=0$.

\subsection{The role of the continuity equation}
It is also worth pointing out that the standard proof of the theorem for
STDCDFT\cite{DiVentra2007}, and --as we have also shown before-- of the theorem of standard
TD-CDFT,\cite{Vignale2004} does {\it not} require that the particle density in
the auxiliary system is equal to the particle density in the original one.
Indeed, in the standard TD-CDFT proof one uses the standard continuity equation
to {\it infer} the equality between the two densities. In our
proof\cite{DiVentra2007}, we postulated it from the uniqueness of the solution
of the non-linear continuity Eq. (\ref{continuity}), with $\mathcal{F}_B(r,t)$
a functional of the current density. In those proofs one only needs to
determine the $n$-th time derivative of all quantities, and only the $(n+1)$-th
time derivative of the vector potential. The $n$-th time derivative of the
particle density is obtained from Eq. (\ref{continuity}) (or, in a closed
system from the same equation with $\mathcal{F}_B=0$). That equation, however,
does not contain any $(n+1)$-th time derivative. Therefore, the equation for
the vector potential in the auxiliary system we use in our
proof\cite{DiVentra2007} is still a recursive relation, with a unique solution
provided the initial conditions. (See section \ref{standardtdcdft} for more details.) The theorem then guarantees the one-to-one
correspondence between external vector potential and ensemble-averaged current
density, leaving open the (possibly quite difficult) task of obtaining the
ensemble-averaged density from the current density, when the continuity
equation is given by Eq. (\ref{continuity}).

This one-to-one mapping, implies that $\mathcal F_B$ is a functional of the
current density $\vec j(r,t)$, the initial conditions and the coupling between
the system and the environment. This ultimately implies that-- by the knowledge
of this functional--we could uniquely determine the single-particle density.
This is however a common situation with DFT. The theorems of DFT offer a solid
foundation for certain calculations by guaranteeing that one quantity can be
exactly obtained from a {\it doppelganger} system. The other physical
quantities have to be derived from the first one, usually a difficult--if not
hopeless--task.

We now want to critically examine Theorem 3 in Ref.~\onlinecite{Yuen-Zhou2009a},
where the main result of that paper is presented: it is possible to construct a
closed non-interacting quantum system that mimics the dynamics of $n(r,t)$ and
$\vec{j}(r,t)$ of the real open interacting system. In the following, we will
assume that $\mathcal{F}_B(r,t)$ could be written as the divergence of $e
n(r,t)\vec C(r,t)/m$ as in Eq. (\ref{leakage}). We want to point out that, one can in
fact find {\it many} (possibly infinite) closed non-interacting quantum systems
that reproduce the dynamics of the exact current and particle densities. One
can show that this is {\it not} in contrast with the general theory of TDCDFT.
Indeed, we can show that the {\it total} current in the closed KS system is
uniquely determined by the total vector potential, given by the sum of the KS
vector potential and $\vec{C}(r,t)$. However, the particle and the
current density $\vec j(r,t)$ do not share this property, and we can find
infinitely many closed non-interacting KS systems that reproduce these
quantities. Indeed, in the proof of the Theorem 3 in Ref.~\onlinecite{Yuen-Zhou2009a}
it is stated that the vector
\begin{equation}
	\vec{C}(r,t)=-\frac{m}{e n(r,t)} \int d^3r \left(\frac{\partial n(r,t)}{\partial
t}+\vec{\nabla}\cdot \vec{j}(r,t)\right)\label{VecC}
\end{equation}
can be uniquely determined by Eqs. (\ref{continuity}) and (\ref{leakage}) once
a boundary condition in space that fixes an arbitrary function of time is
assigned.

However, the above equation is not the unique solution. For instance, easily
fulfilling the assigned boundary condition, one can add the curl of an
arbitrary vector, $\vec{g}(r,t)$, to the leakage potential $\vec{C}(r,t)$
and still satisfy Eq. (\ref{VecC}). Indeed, it is easily proven that
$\vec{C}'(r,t)=\vec{C}(r,t) +(\vec{\nabla}\times\vec{g}(r,t))/n(r,t)$ still
satisfies Eqs. (\ref{continuity}) and (\ref{leakage}), if $\vec{C}(r,t)$ satisfies
the same equation. If we now continue with the proof and use $\vec{C}'(r,t)$ we
arrive at a new vector potential $\vec{A}_{KS}'(r,t)$ that gives the same
$n(r,t)$ and $\vec{j}(r,t)$ as the couple $\vec{A}_{KS}(r,t)$ and
$\vec{C}(r,t)$. This ambiguity reflects the fact that we are trying to mimic
the effect of a scalar function - the term ${\cal F}_B(r,t)$ in Eq.
(\ref{continuity}) - with a vector function, $\vec{C}(r,t)$, without imposing
strict boundary conditions (BCs) on $\vec{C}(r,t)$.

A simple solution to the aforementioned problem may appear by setting
$\vec{\nabla} \times \vec{C}(r,t)\equiv 0$. However, it is important
to realize that the imposition of certain boundary conditions on the
dynamics of these quantities has a direct impact on the uniqueness of
the results. For example, assuming that $\vec{C}(r,t)$ reaches a
certain uniform limit when $|r|\to \infty$, might be {\it inconsistent} with
the bath operator acting on the true many-body system. Indeed, certain
bath operators can be strongly non-local in space, effectively
transferring charge from one region of space to another, with the two
arbitrarily far from each other. Therefore, the BCs on $\vec{\nabla}
\times \vec{C}(r,t)$ have not a clear physical origin or relation to
any physical observable. Fixing their value to obtain one solution
appears utterly arbitrary.

Notice that a similar problem appears also in the case of the standard theorem
of TDDFT (see for example Ref.~\onlinecite{vanLeeuwen1999}) where the
additional boundary condition $n(r)\vec{\nabla} \Delta V(r,t)\to 0$ when
$|r|\to \infty$ is added to the proof. However, in this case, while in
principle this condition is arbitrary and one may choose another condition,
this choice is motivated by physical arguments that are valid for a wide range
of systems. Further investigation of the role of this boundary condition on the
validity of the theorem for standard TDDFT has been carried out in Refs.
\onlinecite{Ruggenthaler2009},\onlinecite{Ruggenthaler2011}, and
\onlinecite{Ruggenthaler2012} The same considerations instead do not apply to
all the components of the vector $\vec{C}(r,t)$.

\section{A mapping theorem: from open to closed quantum system}
These ambiguities derive again from the attempt to find a mapping between the
vector potential and the particle and current densities, when the continuity
equation is not valid. In this attempt, one needs to include another scalar
quantity to make that mapping meaningful. The choice of a vector quantity like
$\vec C(r,t)$ in Eq. (\ref{VecC}) is ill-posed and leads to another ambiguity.
On the other hand, if we are to gain any physical insight on the particle
density, we need to obtain reliable approximations of the leakage potential.
Instead, if one accepts that the quantity of interest is the current density,
the leakage potential must be a functional of the current density and therefore
we could solve the continuity equation for the particle density. For this
reason, we can prove the following theorem (see below about some caveats regarding the use of a
density matrix vs. a state vector formulation)

\emph{Theorem A (unique mapping from open to closed system)}: Consider the
dynamics of a many-body system in contact with an external environment. Assume
the evolution of the density matrix is given by Eq. (\ref{master}) with the
full many-body Hamiltonian (\ref{h_manybody}). Then under reasonable physical
assumptions, there exists a {\it
closed} non-interacting auxiliary many-body Kohn-Sham system which starting from given initial conditions for the state of
the many-body open system, evolves according to
\begin{equation}
	i\partial_t |\Psi_{KS}(t)\rangle=\hat H_{KS}(t) |\Psi_{KS}(t)\rangle
\end{equation}
with Hamiltonian
\begin{equation}
	\hat H_{KS}(t)=\sum_i\left(\frac{\hat{\vec p}_i+e\vec A_{KS}(\hat r_i,t)}{2m}\right)^2
\end{equation}
and reproduces the dynamics of the current density of the original many-body {\it open} system,
\begin{equation}
	\vec j_{KS}(r,t)=\vec j(r,t).
\end{equation}
The proof of this theorem is identical to those already present in the
literature,\cite{DiVentra2007,DAgosta2008a,Biele2012} and with little
modifications to the one we propose in Sec. \ref{standardtdcdft}, we are
therefore not reproducing it here. The
only difference is that the current density of the original many-body open
system follows the dynamics induced by the many-body Hamiltonian and by the
external environment.\cite{DiVentra2007,DAgosta2008a,Biele2012} The
single-particle density can be obtained from Eq. (\ref{continuity}), once the
leakage potential is written as a functional of $\vec j(r,t)$. The uniqueness
of the closed quantum system then follows from the fact that two closed systems
which share the same single-particle current density and initial conditions do
coincide.\cite{Vignale2004}

It is also interesting to note that the possibility of studying the dynamics of
an open system by using a TDCDFT closed system, has been presented in the past.
For example in Ref. \onlinecite{DAgosta2007} the dynamics of a 2D electron gas
coupled with the electrons confined in a 1D quantum well is investigated. The
quantum well, via the Coulomb interaction, acts as an external forcing field
that provides energy to the 2D gas. The dynamics of the latter is then
described by an effective vector potential which contains no reference to the
1D well. One can show that the 2D electron gas will relax to a steady state
regime. A similar analysis has been performed for a 1D electron gas confined in
a quantum well by using the Vignale-Kohn functional for
TDCDFT.\cite{Vignale1996b} In that case, the system relaxes to the ground
state.\cite{Wijewardane2005,DAgosta2006} The KS energy is lost to those degrees
of freedom that the Vignale-Kohn functional does not take into account in
describing the dynamics of the many-body system through some kind of common
variable like the single-particle current density. It should not be surprising
then that in this case the KS energy could be given a physical interpretation
as the maximum work that could be extracted from the system.\cite{DAgosta2006}
More recently, Tokatly\cite{Tokatly2013} has presented a mapping between an
open quantum system, a many-body electron system in contact with the radiation
of a cavity, with a KS closed system for the investigation of the dynamics of
the electrons, together with an alternative proof of the theorem we present in
sec. \ref{standardtdcdft} based on the use of a non-linear Schr\"odinger
equation (see Ref. \onlinecite{Tokatly2013} and references therein).

Finally, we want to comment on a fundamental but important issue. As we have
discussed at length in our previous publications,\cite{DiVentra2007,
DAgosta2008a} and at the beginning of this paper, one has two choices to
describe an open quantum system: in terms of the density matrix or the state
vector. These two formulations are generally equivalent after the observables calculated with the state vector are averaged over the many replicas of the system built to reproduce the mixed state dynamics proper of an open quantum system, being the state vector
formulation an unraveling of the density matrix one. However, the density
matrix approach suffers from two drawbacks which do not make it a solid
starting point for a formulation of DFT for open quantum systems. This is due
to both the possible loss of positivity of the density matrix if an equation of
motion of such quantity is employed with the Hamiltonian and/or bath
operator(s) dependent on time,\cite{Ford2005} and the fact that the KS
Hamiltonian does depend on internal degrees of freedom. Starting from the
master equation formulation of the same problem, one needs to exclude from the
outset the possibility that the Hamiltonian of any auxiliary system with
different interaction potential (and hence the KS Hamiltonian) depends on the
internal degrees of freedom. Otherwise, for such a system no closed
density-matrix equation can be obtained. In other words, one needs to start
from an hypothesis that constitutes part of the final thesis. It is only when
one starts from a stochastic Schr\"odinger equation for the {\em state vector}
that one can prove that the {\em exact} KS Hamiltonian depends only on the
average current density \cite{DiVentra2007,
DAgosta2008a}. In view of this criticism, we can easily reformulate
the theorem A by starting from the stochastic Schr\"odinger
equation as we have done in Ref. \onlinecite{DiVentra2007}.

\section{Conclusions}
In conclusion, we hope we have clarified many misunderstandings regarding a DFT
for open quantum systems and its theoretical foundations. We have also shown
some of the pitfalls of recent theories that have been advanced to improve on
such theory. Such pitfalls originate from a simple, but yet not fully
appreciated point in the DFT community, namely that one cannot map vector
potentials with single-particle densities, a trivial consequence of the fact
that there is no one-to-one correspondence between a vector and a scalar.

\acknowledgments
RD'A acknowledges the support of the Thomas Young Centre in London at Imperial College and King's College London, the support from Grupos Consolidados UPV/EHU del Gobierno Vasco
(IT-319-07), and ACI-Promociona (ACI2009-1036), and the financial support of the
CONSOLIDER-INGENIO 2010 ``NanoTherm" (CSD2010-00044). MD acknowledges financial
support from the Department of Energy grant DE-FG02-05ER46204.

\bibliography{library}

\begin{thebibliography}{42}
\expandafter\ifx\csname natexlab\endcsname\relax\def\natexlab#1{#1}\fi
\expandafter\ifx\csname bibnamefont\endcsname\relax
  \def\bibnamefont#1{#1}\fi
\expandafter\ifx\csname bibfnamefont\endcsname\relax
  \def\bibfnamefont#1{#1}\fi
\expandafter\ifx\csname citenamefont\endcsname\relax
  \def\citenamefont#1{#1}\fi
\expandafter\ifx\csname url\endcsname\relax
  \def\url#1{\texttt{#1}}\fi
\expandafter\ifx\csname urlprefix\endcsname\relax\def\urlprefix{URL }\fi
\providecommand{\bibinfo}[2]{#2}
\providecommand{\eprint}[2][]{\url{#2}}

\bibitem[{\citenamefont{Zwanzig}(1960)}]{Zwanzig1960}
\bibinfo{author}{\bibfnamefont{R.}~\bibnamefont{Zwanzig}}, \bibinfo{journal}{J.
  Chem. Phys.} \textbf{\bibinfo{volume}{33}}, \bibinfo{pages}{1338}
  (\bibinfo{year}{1960}).

\bibitem[{\citenamefont{Nakajima}(1958)}]{Nakajima1958}
\bibinfo{author}{\bibfnamefont{S.}~\bibnamefont{Nakajima}},
  \bibinfo{journal}{Prog. Theor. Phys.} \textbf{\bibinfo{volume}{20}},
  \bibinfo{pages}{948} (\bibinfo{year}{1958}).

\bibitem[{\citenamefont{Gisin}(1981)}]{Gisin1981}
\bibinfo{author}{\bibfnamefont{N.}~\bibnamefont{Gisin}},
  \bibinfo{journal}{Journal of Physics A: Mathematical and Theoretical}
  \textbf{\bibinfo{volume}{14}}, \bibinfo{pages}{2259} (\bibinfo{year}{1981}),
  \urlprefix\url{http://iopscience.iop.org/0305-4470/14/9/021/pdf/0305-4470\_14\_9\_021.pdf}.

\bibitem[{\citenamefont{Strunz}(1996)}]{Strunz1996}
\bibinfo{author}{\bibfnamefont{W.~T.} \bibnamefont{Strunz}},
  \bibinfo{journal}{Phys. Rev. A} \textbf{\bibinfo{volume}{54}},
  \bibinfo{pages}{2664} (\bibinfo{year}{1996}), ISSN \bibinfo{issn}{1050-2947},
  \urlprefix\url{http://www.ncbi.nlm.nih.gov/pubmed/9913775}.

\bibitem[{\citenamefont{Hohenberg and Kohn}(1964)}]{Hohenberg1964}
\bibinfo{author}{\bibfnamefont{P.}~\bibnamefont{Hohenberg}} \bibnamefont{and}
  \bibinfo{author}{\bibfnamefont{W.}~\bibnamefont{Kohn}},
  \bibinfo{journal}{Phys. Rev.} \textbf{\bibinfo{volume}{136}},
  \bibinfo{pages}{B864} (\bibinfo{year}{1964}), ISSN \bibinfo{issn}{0031-899X},
  \urlprefix\url{http://link.aps.org/doi/10.1103/PhysRev.136.B864}.

\bibitem[{\citenamefont{Kohn and Sham}(1965)}]{Kohn1965}
\bibinfo{author}{\bibfnamefont{W.}~\bibnamefont{Kohn}} \bibnamefont{and}
  \bibinfo{author}{\bibfnamefont{L.~J.} \bibnamefont{Sham}},
  \bibinfo{journal}{Phys. Rev.} \textbf{\bibinfo{volume}{140}},
  \bibinfo{pages}{A1133} (\bibinfo{year}{1965}), ISSN
  \bibinfo{issn}{0031-899X},
  \urlprefix\url{http://link.aps.org/doi/10.1103/PhysRev.140.A1133}.

\bibitem[{\citenamefont{Dreizler and Gross}(1990)}]{Dreizler1990}
\bibinfo{author}{\bibfnamefont{R.~M.} \bibnamefont{Dreizler}} \bibnamefont{and}
  \bibinfo{author}{\bibfnamefont{E.~K.~U.} \bibnamefont{Gross}},
  \emph{\bibinfo{title}{{Density Functional Theory}}}
  (\bibinfo{publisher}{Springer-Verlag}, \bibinfo{address}{Heidelberg},
  \bibinfo{year}{1990}).

\bibitem[{\citenamefont{L\"{o}wdin and Trickey}(1990)}]{Trickey1990}
\bibinfo{author}{\bibfnamefont{P.~O.} \bibnamefont{L\"{o}wdin}}
  \bibnamefont{and} \bibinfo{author}{\bibfnamefont{S.~B.}
  \bibnamefont{Trickey}}, \emph{\bibinfo{title}{{Density functional theory of
  many-fermion systems}}}, vol.~\bibinfo{volume}{21} of
  \emph{\bibinfo{series}{Advances in Quantum Chemistry}}
  (\bibinfo{publisher}{Academic Press}, \bibinfo{year}{1990}),
  \urlprefix\url{http://books.google.com/books?hl=en\&amp;lr=\&amp;id=Kqd0Exouk3IC\&amp;oi=fnd\&amp;pg=PR12\&amp;dq=Density+Functional+Theory+of+Many-Fermion+Systems\&amp;ots=tMb0MsKI\_y\&amp;sig=5hfVf8MbGFhOjI2yqguQ\_sqPDsA}.

\bibitem[{\citenamefont{Nalewajski}(1996)}]{Nalewajski1996}
\bibinfo{editor}{\bibfnamefont{R.}~\bibnamefont{Nalewajski}}, ed.,
  \emph{\bibinfo{title}{{Density Functional Theory}}}, vol.
  \bibinfo{volume}{181} of \emph{\bibinfo{series}{Topics in Current Chemistry}}
  (\bibinfo{publisher}{Springer Berlin}, \bibinfo{address}{Berlin},
  \bibinfo{year}{1996}), ISBN \bibinfo{isbn}{978-3-540-61131-8}.

\bibitem[{\citenamefont{Giuliani and Vignale}(2005)}]{Giulianivignale}
\bibinfo{author}{\bibfnamefont{G.~F.} \bibnamefont{Giuliani}} \bibnamefont{and}
  \bibinfo{author}{\bibfnamefont{G.}~\bibnamefont{Vignale}},
  \emph{\bibinfo{title}{{Quantum Theory of the Electron Liquid}}}
  (\bibinfo{publisher}{Cambridge}, \bibinfo{address}{Cambridge, UK},
  \bibinfo{year}{2005}), \bibinfo{edition}{1st} ed.

\bibitem[{\citenamefont{{Di Ventra}}(2008)}]{DiVentra2008}
\bibinfo{author}{\bibfnamefont{M.}~\bibnamefont{{Di Ventra}}},
  \emph{\bibinfo{title}{{Electrical transport in nanoscale systems}}}
  (\bibinfo{publisher}{Cambridge University Press}, \bibinfo{address}{New
  York}, \bibinfo{year}{2008}), \bibinfo{edition}{1st} ed., ISBN
  \bibinfo{isbn}{978-0-521-89634-4}.

\bibitem[{\citenamefont{Marques et~al.}(2006)\citenamefont{Marques, Ullrich,
  Rubio, Nogueira, Burke, and Gross}}]{Marques2006}
\bibinfo{editor}{\bibfnamefont{M.~A.~L.} \bibnamefont{Marques}},
  \bibinfo{editor}{\bibfnamefont{C.~A.} \bibnamefont{Ullrich}},
  \bibinfo{editor}{\bibfnamefont{A.}~\bibnamefont{Rubio}},
  \bibinfo{editor}{\bibfnamefont{F.}~\bibnamefont{Nogueira}},
  \bibinfo{editor}{\bibfnamefont{K.}~\bibnamefont{Burke}}, \bibnamefont{and}
  \bibinfo{editor}{\bibfnamefont{E.~K.~U.} \bibnamefont{Gross}}, eds.,
  \emph{\bibinfo{title}{{Time-Dependent Density Functional Theory}}}, vol.
  \bibinfo{volume}{706} of \emph{\bibinfo{series}{Lecture notes in Physics}}
  (\bibinfo{publisher}{Springer}, \bibinfo{address}{Berlin},
  \bibinfo{year}{2006}), ISBN \bibinfo{isbn}{3-540-35422-0}.

\bibitem[{\citenamefont{Burke et~al.}(2005)\citenamefont{Burke, Car, and
  Gebauer}}]{Burke2005}
\bibinfo{author}{\bibfnamefont{K.}~\bibnamefont{Burke}},
  \bibinfo{author}{\bibfnamefont{R.}~\bibnamefont{Car}}, \bibnamefont{and}
  \bibinfo{author}{\bibfnamefont{R.}~\bibnamefont{Gebauer}},
  \bibinfo{journal}{Phys. Rev. Lett.} \textbf{\bibinfo{volume}{94}},
  \bibinfo{pages}{146803} (\bibinfo{year}{2005}).

\bibitem[{\citenamefont{{Di Ventra} and D'Agosta}(2007)}]{DiVentra2007}
\bibinfo{author}{\bibfnamefont{M.}~\bibnamefont{{Di Ventra}}} \bibnamefont{and}
  \bibinfo{author}{\bibfnamefont{R.}~\bibnamefont{D'Agosta}},
  \bibinfo{journal}{Phys. Rev. Lett.} \textbf{\bibinfo{volume}{98}},
  \bibinfo{pages}{226403} (\bibinfo{year}{2007}).

\bibitem[{\citenamefont{D'Agosta and {Di Ventra}}(2008)}]{DAgosta2008a}
\bibinfo{author}{\bibfnamefont{R.}~\bibnamefont{D'Agosta}} \bibnamefont{and}
  \bibinfo{author}{\bibfnamefont{M.}~\bibnamefont{{Di Ventra}}},
  \bibinfo{journal}{Phys. Rev. B} \textbf{\bibinfo{volume}{78}},
  \bibinfo{pages}{165105} (\bibinfo{year}{2008}).

\bibitem[{\citenamefont{Appel and {Di Ventra}}(2009)}]{Appel2009}
\bibinfo{author}{\bibfnamefont{H.}~\bibnamefont{Appel}} \bibnamefont{and}
  \bibinfo{author}{\bibfnamefont{M.}~\bibnamefont{{Di Ventra}}},
  \bibinfo{journal}{Phys. Rev. B} \textbf{\bibinfo{volume}{80}},
  \bibinfo{pages}{212303} (\bibinfo{year}{2009}), ISSN
  \bibinfo{issn}{1098-0121},
  \urlprefix\url{http://link.aps.org/doi/10.1103/PhysRevB.80.212303}.

\bibitem[{\citenamefont{Yuen-Zhou et~al.}(2009)\citenamefont{Yuen-Zhou,
  Rodr\'{\i}guez-Rosario, and Aspuru-Guzik}}]{Yuen-Zhou2009a}
\bibinfo{author}{\bibfnamefont{J.}~\bibnamefont{Yuen-Zhou}},
  \bibinfo{author}{\bibfnamefont{C.}~\bibnamefont{Rodr\'{\i}guez-Rosario}},
  \bibnamefont{and}
  \bibinfo{author}{\bibfnamefont{A.}~\bibnamefont{Aspuru-Guzik}},
  \bibinfo{journal}{Phys. Chem. Chem. Phys.} \textbf{\bibinfo{volume}{11}},
  \bibinfo{pages}{4509} (\bibinfo{year}{2009}), ISSN \bibinfo{issn}{1463-9076},
  \urlprefix\url{http://www.ncbi.nlm.nih.gov/pubmed/19475169}.

\bibitem[{\citenamefont{Yuen-Zhou et~al.}(2010)\citenamefont{Yuen-Zhou, Tempel,
  Rodr\'{\i}guez-Rosario, and Aspuru-Guzik}}]{Yuen-Zhou2010}
\bibinfo{author}{\bibfnamefont{J.}~\bibnamefont{Yuen-Zhou}},
  \bibinfo{author}{\bibfnamefont{D.~G.} \bibnamefont{Tempel}},
  \bibinfo{author}{\bibfnamefont{C.~A.} \bibnamefont{Rodr\'{\i}guez-Rosario}},
  \bibnamefont{and}
  \bibinfo{author}{\bibfnamefont{A.}~\bibnamefont{Aspuru-Guzik}},
  \bibinfo{journal}{Phys. Rev. Lett.} \textbf{\bibinfo{volume}{104}},
  \bibinfo{pages}{043001} (\bibinfo{year}{2010}), ISSN
  \bibinfo{issn}{0031-9007},
  \urlprefix\url{http://link.aps.org/doi/10.1103/PhysRevLett.104.043001}.

\bibitem[{\citenamefont{Biele and D'Agosta}(2012)}]{Biele2012}
\bibinfo{author}{\bibfnamefont{R.}~\bibnamefont{Biele}} \bibnamefont{and}
  \bibinfo{author}{\bibfnamefont{R.}~\bibnamefont{D'Agosta}},
  \bibinfo{journal}{J. Phys.: Cond. Matt.} \textbf{\bibinfo{volume}{24}},
  \bibinfo{pages}{273201} (\bibinfo{year}{2012}), ISSN
  \bibinfo{issn}{1361-648X}, \eprint{arXiv:1112.2694v1},
  \urlprefix\url{http://www.ncbi.nlm.nih.gov/pubmed/22713734}.

\bibitem[{\citenamefont{Appel and {Di Ventra}}(2011)}]{Appel2011}
\bibinfo{author}{\bibfnamefont{H.}~\bibnamefont{Appel}} \bibnamefont{and}
  \bibinfo{author}{\bibfnamefont{M.}~\bibnamefont{{Di Ventra}}},
  \bibinfo{journal}{Chem. Phys.} \textbf{\bibinfo{volume}{391}},
  \bibinfo{pages}{27} (\bibinfo{year}{2011}), ISSN \bibinfo{issn}{03010104},
  \eprint{1101.3079},
  \urlprefix\url{http://linkinghub.elsevier.com/retrieve/pii/S0301010411001558}.

\bibitem[{\citenamefont{Pershin et~al.}(2008)\citenamefont{Pershin, Dubi, and
  {Di Ventra}}}]{Pershin2008a}
\bibinfo{author}{\bibfnamefont{Y.~V.} \bibnamefont{Pershin}},
  \bibinfo{author}{\bibfnamefont{Y.}~\bibnamefont{Dubi}}, \bibnamefont{and}
  \bibinfo{author}{\bibfnamefont{M.}~\bibnamefont{{Di Ventra}}},
  \bibinfo{journal}{Phys. Rev. B} \textbf{\bibinfo{volume}{78}},
  \bibinfo{pages}{054302} (\bibinfo{year}{2008}), ISSN
  \bibinfo{issn}{1098-0121},
  \urlprefix\url{http://link.aps.org/doi/10.1103/PhysRevB.78.054302}.

\bibitem[{\citenamefont{Vignale et~al.}(1988)\citenamefont{Vignale, Rasolt, and
  Geldart}}]{Vignale1988}
\bibinfo{author}{\bibfnamefont{G.}~\bibnamefont{Vignale}},
  \bibinfo{author}{\bibfnamefont{M.}~\bibnamefont{Rasolt}}, \bibnamefont{and}
  \bibinfo{author}{\bibfnamefont{D.~J.~W.} \bibnamefont{Geldart}},
  \bibinfo{journal}{Phys. Rev. B} \textbf{\bibinfo{volume}{37}},
  \bibinfo{pages}{2502} (\bibinfo{year}{1988}).

\bibitem[{\citenamefont{Frensley}(1990)}]{Frensley1990}
\bibinfo{author}{\bibfnamefont{W.~R.} \bibnamefont{Frensley}},
  \bibinfo{journal}{Rev. Mod. Phys.} \textbf{\bibinfo{volume}{62}},
  \bibinfo{pages}{745} (\bibinfo{year}{1990}).

\bibitem[{\citenamefont{Gardiner and Zoeller}(2000)}]{Gardiner2000}
\bibinfo{author}{\bibfnamefont{C.~W.} \bibnamefont{Gardiner}} \bibnamefont{and}
  \bibinfo{author}{\bibfnamefont{P.}~\bibnamefont{Zoeller}},
  \emph{\bibinfo{title}{{Quantum Noise}}} (\bibinfo{publisher}{Springer},
  \bibinfo{address}{Berlin}, \bibinfo{year}{2000}), \bibinfo{edition}{2nd} ed.,
  ISBN \bibinfo{isbn}{3-540-66571-4}.

\bibitem[{\citenamefont{van Kampen}(2007)}]{vanKampen}
\bibinfo{author}{\bibfnamefont{N.~G.} \bibnamefont{van Kampen}},
  \emph{\bibinfo{title}{{Stochastic Processes in Physics and Chemistry}}}
  (\bibinfo{publisher}{Elsevier}, \bibinfo{address}{Amsterdam},
  \bibinfo{year}{2007}), \bibinfo{edition}{3rd} ed., ISBN
  \bibinfo{isbn}{978-0-444-52965-7}.

\bibitem[{\citenamefont{Gebauer and Car}(2004)}]{Gebauer2004a}
\bibinfo{author}{\bibfnamefont{R.}~\bibnamefont{Gebauer}} \bibnamefont{and}
  \bibinfo{author}{\bibfnamefont{R.}~\bibnamefont{Car}},
  \bibinfo{journal}{Phys. Rev. Lett.} \textbf{\bibinfo{volume}{93}},
  \bibinfo{pages}{160404} (\bibinfo{year}{2004}), ISSN
  \bibinfo{issn}{0031-9007},
  \urlprefix\url{http://link.aps.org/doi/10.1103/PhysRevLett.93.160404}.

\bibitem[{\citenamefont{Gebauer et~al.}(2005)\citenamefont{Gebauer, Piccinin,
  and Car}}]{Gebauer2005}
\bibinfo{author}{\bibfnamefont{R.}~\bibnamefont{Gebauer}},
  \bibinfo{author}{\bibfnamefont{S.}~\bibnamefont{Piccinin}}, \bibnamefont{and}
  \bibinfo{author}{\bibfnamefont{R.}~\bibnamefont{Car}},
  \bibinfo{journal}{ChemPhysChem} \textbf{\bibinfo{volume}{6}},
  \bibinfo{pages}{1727} (\bibinfo{year}{2005}), ISSN \bibinfo{issn}{1439-4235},
  \urlprefix\url{http://www.ncbi.nlm.nih.gov/pubmed/16144008}.

\bibitem[{\citenamefont{Salmilehto et~al.}(2012)\citenamefont{Salmilehto,
  Solinas, and M\"{o}tt\"{o}nen}}]{Salmilehto2012}
\bibinfo{author}{\bibfnamefont{J.}~\bibnamefont{Salmilehto}},
  \bibinfo{author}{\bibfnamefont{P.}~\bibnamefont{Solinas}}, \bibnamefont{and}
  \bibinfo{author}{\bibfnamefont{M.}~\bibnamefont{M\"{o}tt\"{o}nen}},
  \bibinfo{journal}{Phys. Rev. A} \textbf{\bibinfo{volume}{85}},
  \bibinfo{pages}{032110} (\bibinfo{year}{2012}), ISSN
  \bibinfo{issn}{1050-2947},
  \urlprefix\url{http://link.aps.org/doi/10.1103/PhysRevA.85.032110}.

\bibitem[{\citenamefont{Vignale}(2004)}]{Vignale2004}
\bibinfo{author}{\bibfnamefont{G.}~\bibnamefont{Vignale}},
  \bibinfo{journal}{Phys. Rev. B} \textbf{\bibinfo{volume}{70}},
  \bibinfo{pages}{201102(R)} (\bibinfo{year}{2004}), ISSN
  \bibinfo{issn}{1098-0121},
  \urlprefix\url{http://link.aps.org/doi/10.1103/PhysRevB.70.201102}.

\bibitem[{\citenamefont{D'Agosta and Vignale}(2005)}]{DAgosta2005a}
\bibinfo{author}{\bibfnamefont{R.}~\bibnamefont{D'Agosta}} \bibnamefont{and}
  \bibinfo{author}{\bibfnamefont{G.}~\bibnamefont{Vignale}},
  \bibinfo{journal}{Phys. Rev. B} \textbf{\bibinfo{volume}{71}},
  \bibinfo{pages}{245103} (\bibinfo{year}{2005}), ISSN
  \bibinfo{issn}{1098-0121},
  \urlprefix\url{http://link.aps.org/doi/10.1103/PhysRevB.71.245103}.

\bibitem[{\citenamefont{Runge and Gross}(1984)}]{Runge1984}
\bibinfo{author}{\bibfnamefont{E.}~\bibnamefont{Runge}} \bibnamefont{and}
  \bibinfo{author}{\bibfnamefont{E.~K.~U.} \bibnamefont{Gross}},
  \bibinfo{journal}{Phys. Rev. Lett.} \textbf{\bibinfo{volume}{52}},
  \bibinfo{pages}{997} (\bibinfo{year}{1984}), ISSN \bibinfo{issn}{0031-9007},
  \urlprefix\url{http://link.aps.org/doi/10.1103/PhysRevLett.52.997}.

\bibitem[{\citenamefont{van Leeuwen}(1999)}]{vanLeeuwen1999}
\bibinfo{author}{\bibfnamefont{R.}~\bibnamefont{van Leeuwen}},
  \bibinfo{journal}{Physical Review Letters} \textbf{\bibinfo{volume}{82}},
  \bibinfo{pages}{3863} (\bibinfo{year}{1999}), ISSN \bibinfo{issn}{0031-9007},
  \urlprefix\url{http://link.aps.org/doi/10.1103/PhysRevLett.82.3863}.

\bibitem[{\citenamefont{Ruggenthaler and van Leeuwen}(2011)}]{Ruggenthaler2011}
\bibinfo{author}{\bibfnamefont{M.}~\bibnamefont{Ruggenthaler}}
  \bibnamefont{and} \bibinfo{author}{\bibfnamefont{R.}~\bibnamefont{van
  Leeuwen}}, \bibinfo{journal}{EPL (Europhysics Letters)}
  \textbf{\bibinfo{volume}{95}}, \bibinfo{pages}{13001} (\bibinfo{year}{2011}),
  ISSN \bibinfo{issn}{0295-5075},
  \urlprefix\url{http://stacks.iop.org/0295-5075/95/i=1/a=13001?key=crossref.ff15d8e577186a0fc3ff16060cb4a2fe}.

\bibitem[{\citenamefont{Tokatly}(2011)}]{Tokatly2011}
\bibinfo{author}{\bibfnamefont{I.~V.} \bibnamefont{Tokatly}},
  \bibinfo{journal}{Phys. Rev. B} \textbf{\bibinfo{volume}{83}},
  \bibinfo{pages}{035127} (\bibinfo{year}{2011}), ISSN
  \bibinfo{issn}{1098-0121},
  \urlprefix\url{http://link.aps.org/doi/10.1103/PhysRevB.83.035127}.

\bibitem[{\citenamefont{Ruggenthaler et~al.}(2009)\citenamefont{Ruggenthaler,
  Penz, and Bauer}}]{Ruggenthaler2009}
\bibinfo{author}{\bibfnamefont{M.}~\bibnamefont{Ruggenthaler}},
  \bibinfo{author}{\bibfnamefont{M.}~\bibnamefont{Penz}}, \bibnamefont{and}
  \bibinfo{author}{\bibfnamefont{D.}~\bibnamefont{Bauer}}, \bibinfo{journal}{J.
  Phys. A} \textbf{\bibinfo{volume}{42}}, \bibinfo{pages}{425207}
  (\bibinfo{year}{2009}), ISSN \bibinfo{issn}{1751-8113},
  \urlprefix\url{http://stacks.iop.org/1751-8121/42/i=42/a=425207?key=crossref.72066fdadf96506403285aa756eff495}.

\bibitem[{\citenamefont{Ruggenthaler et~al.}(2012)\citenamefont{Ruggenthaler,
  Giesbertz, Penz, and van Leeuwen}}]{Ruggenthaler2012}
\bibinfo{author}{\bibfnamefont{M.}~\bibnamefont{Ruggenthaler}},
  \bibinfo{author}{\bibfnamefont{K.~J.~H.} \bibnamefont{Giesbertz}},
  \bibinfo{author}{\bibfnamefont{M.}~\bibnamefont{Penz}}, \bibnamefont{and}
  \bibinfo{author}{\bibfnamefont{R.}~\bibnamefont{van Leeuwen}},
  \bibinfo{journal}{Phys. Rev, A} \textbf{\bibinfo{volume}{85}},
  \bibinfo{pages}{052504} (\bibinfo{year}{2012}), ISSN
  \bibinfo{issn}{1050-2947},
  \urlprefix\url{http://link.aps.org/doi/10.1103/PhysRevA.85.052504}.

\bibitem[{\citenamefont{D’Agosta et~al.}(2007)\citenamefont{D’Agosta, {Di
  Ventra}, and Vignale}}]{DAgosta2007}
\bibinfo{author}{\bibfnamefont{R.}~\bibnamefont{D’Agosta}},
  \bibinfo{author}{\bibfnamefont{M.}~\bibnamefont{{Di Ventra}}},
  \bibnamefont{and} \bibinfo{author}{\bibfnamefont{G.}~\bibnamefont{Vignale}},
  \bibinfo{journal}{Phys. Rev. B} \textbf{\bibinfo{volume}{76}},
  \bibinfo{pages}{035320} (\bibinfo{year}{2007}), ISSN
  \bibinfo{issn}{1098-0121},
  \urlprefix\url{http://link.aps.org/doi/10.1103/PhysRevB.76.035320}.

\bibitem[{\citenamefont{Vignale and Kohn}(1996)}]{Vignale1996b}
\bibinfo{author}{\bibfnamefont{G.}~\bibnamefont{Vignale}} \bibnamefont{and}
  \bibinfo{author}{\bibfnamefont{W.}~\bibnamefont{Kohn}},
  \bibinfo{journal}{Phys. Rev. Lett.} \textbf{\bibinfo{volume}{77}},
  \bibinfo{pages}{2037} (\bibinfo{year}{1996}), ISSN \bibinfo{issn}{0031-9007},
  \urlprefix\url{http://link.aps.org/doi/10.1103/PhysRevLett.77.2037}.

\bibitem[{\citenamefont{Wijewardane and Ullrich}(2005)}]{Wijewardane2005}
\bibinfo{author}{\bibfnamefont{H.~O.} \bibnamefont{Wijewardane}}
  \bibnamefont{and} \bibinfo{author}{\bibfnamefont{C.~A.}
  \bibnamefont{Ullrich}}, \bibinfo{journal}{Phys. Rev. Lett.}
  \textbf{\bibinfo{volume}{95}}, \bibinfo{pages}{86401} (\bibinfo{year}{2005}).

\bibitem[{\citenamefont{D'Agosta and Vignale}(2006)}]{DAgosta2006}
\bibinfo{author}{\bibfnamefont{R.}~\bibnamefont{D'Agosta}} \bibnamefont{and}
  \bibinfo{author}{\bibfnamefont{G.}~\bibnamefont{Vignale}},
  \bibinfo{journal}{Phys. Rev. Lett.} \textbf{\bibinfo{volume}{96}},
  \bibinfo{pages}{016405} (\bibinfo{year}{2006}), ISSN
  \bibinfo{issn}{0031-9007}.

\bibitem[{\citenamefont{Tokatly}(2013)}]{Tokatly2013}
\bibinfo{author}{\bibfnamefont{I.~V.} \bibnamefont{Tokatly}},
  \bibinfo{journal}{arXiv preprint arXiv:1303.1947}  (\bibinfo{year}{2013}),
  \eprint{arXiv:1303.1947v1}, \urlprefix\url{http://arxiv.org/abs/1303.1947}.

\bibitem[{\citenamefont{Ford and O'Connell}(2005)}]{Ford2005}
\bibinfo{author}{\bibfnamefont{G.~W.} \bibnamefont{Ford}} \bibnamefont{and}
  \bibinfo{author}{\bibfnamefont{R.~F.} \bibnamefont{O'Connell}},
  \bibinfo{journal}{Ann. Phys.} \textbf{\bibinfo{volume}{319}},
  \bibinfo{pages}{348} (\bibinfo{year}{2005}).

\end{thebibliography}
\end{document}